\documentclass[conference]{IEEEtran}
\IEEEoverridecommandlockouts
\usepackage{cite}
\usepackage{amsmath,amssymb,amsfonts}
\usepackage{algorithmic}
\usepackage{graphicx}
\usepackage{textcomp}
\usepackage{xcolor}
\def\BibTeX{{\rm B\kern-.05em{\sc i\kern-.025em b}\kern-.08em
    T\kern-.1667em\lower.7ex\hbox{E}\kern-.125emX}}
\begin{document}

\title{Interpretation of the principles and implementation of FreeRider: A tutorial
}

\author{\IEEEauthorblockN{1\textsuperscript{st} Chenming Zhang}
\IEEEauthorblockA{\textit{University of Technology and Science of China} \\
\textit{Course Parallel and Distributed Computing} \\
Anhui Hefei, China \\
zhangchm@mail.ustc.edu.cn}
}

\maketitle

\begin{abstract}
We aims to provide an interpretation of the the design background, motivation, and key innovations of FreeRider\cite{c0}.
The technique utilized by FreeRider enables tags to transform codewords present in commodity signals into another valid ones from the same codebook during reflection.
As a result, the backscattered signal remains valid as commodity radios such as ZigBee or Bluetooth.
By using commodity radios, FreeRider's backscatter system addresses the issue of specialized hardware requirements,
thereby expanding its potential applications across various fields such as smart homes, healthcare, and industrial environments.
\end{abstract}


\section{Introduction}
Backscatter communication is a wireless link that allows Internet-of-Things devices to connect while consuming ultra-low power.
Radio Frequency Identification (RFID) is a simple yet effective technology used by businesses, governments, and individuals alike for various purposes such as
inventory management, tracking assets, monitoring production processes, and security applications\cite{c1, c2, c3}.
At its core, RFID consists of two key components: a reader and a tag.
Tags contain a unique identifier or code that is transmitted wirelessly to the reader when the tag comes within range.
The reader then captures this signal and decodes the information stored in the tag.
Compared to active transmissions, passive RFID requires no internal power source and only consumes microwatts of power during data transmission because they use the received electromagnetic waves to recharge, then send messages or modify stored data.

\begin{figure}[htbp]
\centerline{\includegraphics{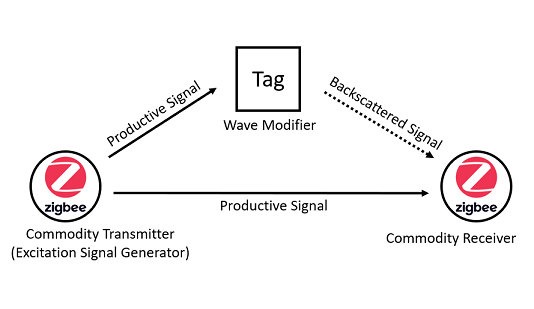}}
\caption{Key components of a backscatter system}
\label{backscatter_components}
\end{figure}

As an extension to RFID, the backscatter system consists of a separate excitation signal generator and receiver shown in Fig.~\ref{backscatter_components}.
Backscatter communication enables battery-free sensors to transmit data using existing commodity radios for they passively reflect and modify wireless signals without any decoding and codeword translation operations to embed information.
The advantage of commodity radio backscatter technology over RFID is that it reduces the complexity and cost of deploying battery-free sensors as data from their tags can be read using existing commodity infrastructures rather than requiring specialized exciters and readers.
Therefore, the potential for wider deployment of this technology exists as it leverages existing infrastructure, reducing costs and complexity.

In order to avoid reader exclusivity, FreeRider\cite{c0} has designed and implemented a backscatter system that can communicate using commodity signals that are not fully dedicated to backscatter devices.
In other words, tags don't need to generate excitation signal while transmitters (gateway or terminal) can directly use the radio communicating with a non-backscattering device as the excitation signal for backscattering.

In the following section, we will look in detail at why FreeRider chose its technical solution, starting with the technical background.
Based on its solution, we perform a brief verification test of the decoding on the receiving end. 
We will also describe what practical values FreeRider achieves and what their future challenges are.

\section{Design Backgrounds and Goals}

While designing a new backscatter system that works on commodity radios, there are aspects to be considered on all three components:
the excitation signal generator, the tag and the receiver.

\subsection{As Exciter}
There have been previous attempts to move away from the specificity of RFID (or backscatter) readers.
Passive WiFi\cite{passive_wifi} tags achieve communication by absorbing and reflecting existing WiFi signals, which makes passive WiFi tags ideal for battery-free IoT devices.
Though passive WiFi shows the compatibility with existing wireless networks, the transmitters have to provide the necessary empty radio strength for communication to take place.
This is a feature not available on commodity WiFi access points or terminals as the 802.11 protocol stack performs whitening operations when sending data.

It has been shown that the performance of IoT devices is already generally affected by ISM band congestion\cite{iot_crowd}.
Due to the limitations of the communication principle in some low-power implementations of backscattering technology,
the tags send data at a lower rate on the carrier waves than its normally encoded data.
This implies light amounts of data from the detector, which also generates a relatively heavy channel load.
Regardless of whether it is a backscatter device or not, we don't expect new technologies to continue to increase the congestion in the ISM band.

The disadvantage of traditional backscatter technology (such as RFID) is that it is limited by distance and requires a dedicated RF signal.
It requires the reader to generate and send an RF signal, then it is expected that the label would receive, obtain energy from induced currents, and reflect the radio wave back to the reader.
The wireless radio fades during the send-and-turn process.
As a result, there is a loss both in the path and at the tag, and the effective communication distance is only several meters\cite{m10,rfid_decay}.
However, without considering the design of the tag internals, it is not necessary for the transmitter to play the role of the receiver\cite{bs_range}.

One key advantage of RFID technology is
its ability to track and manage large numbers of items quickly and accurately.
It enables businesses to automate their inventory processes and track the location of items in real-time, reducing manual errors and labor costs.
In addition, RFID also provides enhanced security measures by accurately identifying and tracking the movement of items, preventing theft and unauthorized access.
We hope that the new backscatter system technology will retain this key advantage.

For these considerations above, on the excitation signal generator (or transmitter) side, we would like to exlore the possibility that:
\begin{itemize}
\item Carrier waves available for backscattering are also able to communicate productively with existing non-backscatter clients.
This reduces the level of communication congestion. 
\item Separating the deployment of the excitation side and the receiver side, thus increasing the flexibility of the device and bringing about a potential increase in communication distance.
\item Sending control message to tags that tags can recognise them in a low-power state. MAC protocol gives the backscatter system the ability to coordinate dozens or more tags.
\end{itemize}

\subsection{As Tag}\label{as_tag}
In general, an IoT system consists of a number of low-energy and power-rich devices, and so does backscatter systems.
A bottleneck in the development and popularity of the Internet of Things is the issue of energy for sensors\cite{energy_tag}.
Most commodity IoT sensors require a stored power supply, such as battery or using solar power.
They are not durable or reliable enough if there is a lack of manual maintenance.

Power consumption can be reduced if the tag does not need to generate its own carrier waves. We can start by looking at how passive RFID tags reflect signals.
RFID readers transmit a continuous waveform as carrier and demodulate the reflected signal,
while tags do not contain any RF Generator.
The tag converts a portion of the signal into energy to maintain the integrated circuit running at the tag,
then reflects another portion of the signal to the reader.
The tag modulates the message onto the carrier sine wave by switching the impedance of the antenna.

\begin{figure}[htbp]
\centerline{\includegraphics{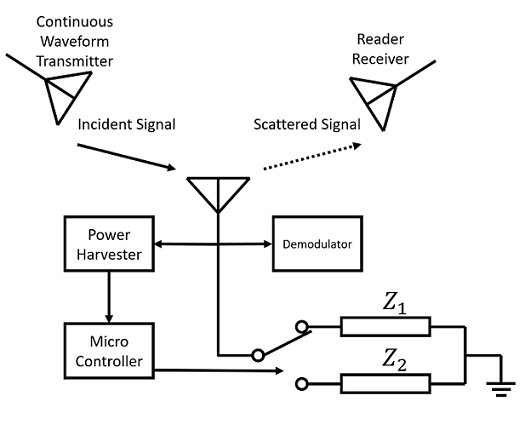}}
\caption{Passive RFID tag structure for reflecting signals}
\label{rfid_tag}
\end{figure}

The backscattered waveform received at the reader is influenced by two relatively independent factors: the antenna structure and the antenna impedance.
Geometrical and material characteristics of the antenna do affect the pattern of scattered signal, but can be detached from the effect of the antenna impedance.
By switching the impedance of the antenna, the tag will send backscattered signal with difference phase and amplitude.
The reader demodulates the reflected signal to recover the information from the tag.
As shown in Fig.~\ref{rfid_tag}, the tag embeds information into the incident continuous waveform by switching its inner impedance between several impedance states.

\begin{figure}[htbp]
\centerline{\includegraphics{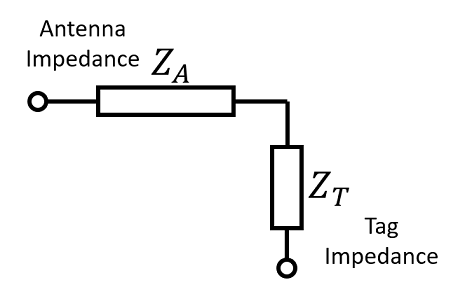}}
\caption{Equivalent RLC circuit for the tag antenna impedance}
\label{thevenin}
\end{figure}

To understand how the impedance affects the backscattered signal,
the antenna structure of the tag can be transformed into an equivalent circuit according to Thevenin's theorem,
as shown in Fig.~\ref{thevenin}.

The impedance of the tag antenna includes the antenna impedance $Z_A = R_A + j X_A$
and the tag impedance $Z_T = R_T + j X_T$ which is controllable for the integrated circuit.
The reflection coefficient is defined as $\Gamma = \frac{Z_T - Z_A^*}{Z_T + Z_A}$, which determines the relationship between the incidence field and scattering field\cite{antenna_property}. 
Maximum transmission strength ($\Gamma = 0$) can be achieved by matching the tag impedance to the antenna ($Z_T = Z_A^*$).
In Fig.~\ref{rfid_tag}, the tag enables the binary modulation of analogue signals by switching the tag impedance between $Z_T = Z_1$ and $Z_T = Z_2$.
We can also accomplish higher order modulation by setting a larger set of tag impedance values.
Radio waves carry information with nothing more than amplitude, frequency and phase.
The switching approach consumes ultra low energy, and it is ideal for application on other commodity radios.

In order to design backscattering tags on commodity radios, we want the backscatter modulation transforms one valid codeword to another valid one.
If the standard waveform carries bits by amplitude, a tag can modify its amplitude, and this is also true for frequency and phase.
The ultimate aim of this is to enable the commodity receivers to decode backscattered bits, rather than forced to equip bespoke ends.
The specific ways in signal modification is described in section \ref{sec3}--\ref{sec4}.

Another step that generates significant energy consumption is the demodulation and decoding of the received signal.
The tag only sends data for a short time and its communication module is mostly dormant.
However, for reasons of functionality and fault tolerance, it is necessary to coordinate with the backscatter controller when to send sensor data.
This requires the tag to keep listening for control messages on the target channel all the time.

Fortunately, tags can get control message from the coordinator without demodulation and decoding.
FreeRider passes control messages to tags via the envelope detector, which detects the length of a packet.
An envelop detector consumes much less power than a demodulator\cite{c0}.
\textit{It is often fisible for the coordinator or exciter to vary the packet length somewhat freely during productive communications,
as most commodity radio protocols support sending data in segments and reassembling it,
and the length of physical layer packets is not visible to application layer data.}

Overall, the tag plays a low-energy role, then there are a number of ways in which we can reduce the power consumption during communications:
\begin{itemize}
\item Eliminates the demand to generate carriers at the tag. This is achieved by reflecting existing commodity waveforms using RFID-like methods.
\item Eliminates the need for demodulation or decoding at the tag. This is achieved by transmitting control messages in packet lengths rather than modulated codewords.
\end{itemize}

\subsection{As Receiver}
The precursor work has given a solution to improve the effective communication distance between the tag and the receiver
by deploying the excitation generator and the receiver separately\cite{increased_range}.
Because power-rich receivers for commodity signals are often numerous in the environment,
it is beneficial to make full use of them to improve the success rate of backscatter communications.

If the packet length modulation mentioned in section \ref{as_tag} is used,
the coordinator has to be specialized for changing packet lengths, but the receiver does not.
Packets with packet length modulation can be aggregated normally by non-backscatter devices.

\begin{figure}[htbp]
\centerline{\includegraphics{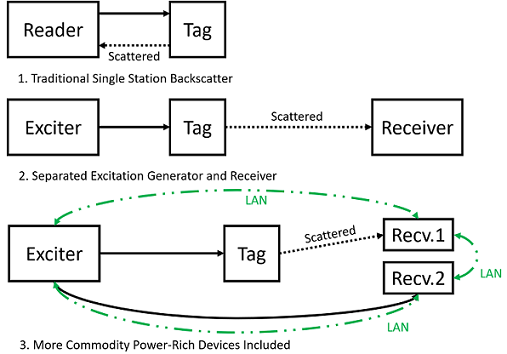}}
\caption{(1) RFID style backscatter system. (2) Deploying the excitation generator and the receiver separately\cite{increased_range}. (3) FreeRider Design\cite{c0}.}
\label{diff}
\end{figure}

Ideally, by getting error reports in the received data packets from physical layer,
any receivers in the local area network (LAN) can co-operate in obtaining the embedded data from the tag, as shown in Fig.~\ref{diff}.

\begin{figure}[htbp]
\centerline{\includegraphics{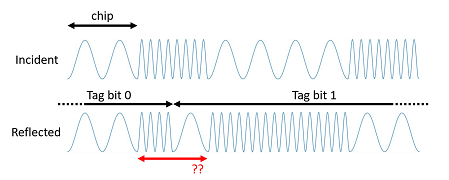}}
\caption{Bad symbol occurs in backscattered signals. Using FSK as an illustration.}
\label{damage}
\end{figure}

According to the tag design discussed in section \ref{as_tag},
since tags are not aware of what information is encoded in the excitation signal,
and unable to distinguish the temporal boundaries of the modulated symbols,
some destructions of symbol waveform structure is inevitable in the received backscattered signal, as shown in Fig.~\ref{damage}.
In order for commodity receivers be able to decode the tag data,
the backscattered signal should contain at least a majority of valid codewords.
We can achieve this by specifying the tag to embed data bits at a lower rate.

The exact time of arrival of the tag bits is not known at the receiver,
creating a greater overhead for tag data demodulation.
Nevertheless, these complexities lead to a simplification on tags,
while they are not a problem for power-rich devices.

If backscatter is performed on a commodity radio, we should take advantage of these following conveniences:

\begin{itemize}
\item A receiver does not necessarily need to take on the task of coordinating tags. Separating them from the exciter or coordinator is an option.
\item Power-rich commodity devices in dispersed physical locations that can collaborate to obtain tag data over a LAN.
Any commodity device in the environment can become a receiver with minor software changes.
\item Power-rich receivers are responsible for the simplicity of tags. The receiver should be able to decode tag bits correctly, even if the tag produces some bad symbols.
\end{itemize}

\section{Productive Carrier Waveforms}\label{sec3}

We have to understand how productive commodity waveforms are modulated and demodulated before modifying them with tag bits.
Our goal is that the signal after tag modulation still contains the majority of valid codewords of the same protocol,
allowing non-backscatter devices to recognise these packets.

\subsection{Amplitude Shift Keying}

A tag can implement ASK on an existing continuous waveform in a simple and low-power way.
For example, a passive RFID tag, described in section \ref{as_tag}, modulates the amplitude by switching the antenna impedance.
By doing so, a continuous waveform will be converted into two different amplitudes before it is correctly decoded by the reader.
The carrier amplitude is constant, so different modification in amplitude represent the corresponding codewords.

However, if we want to use a signal that has been amplitude-modulated as a carrier,
remember that there is no way for a tag to know exactly what each symbol in the carrier contains.
Any ASK standard modulates the signal to a finite number of different amplitudes.
Once the reflection coefficient $\Gamma = \frac{Z_T - Z_A^*}{Z_T + Z_A}$ varies,
it derives at least one additional different amplitude unless it makes the amplitude zero and it represents a valid codeword.
Non-zero amplitude modification by tags either produce bad codewords or is ignored as an attenuation at the receiver.

A productive signal would never be a continuous waveform,
and we do not have an effective means of performing expected non-zero amplitude alternation.
It is preferable to avoid imposing amplitude modulation at the tag
if zero amplitude is not a legal symbol in the original transmission.

\subsection{Frequency Shift Keying}\label{sec_fsk}

In binary frequency shift keying, a bit `1' is transmitted at a higher frequency $f_1$, while a bit `0' is transmitted at a lower frequency $f_0$.
At a frequency interval $\Delta f = f_1 - f_0$, The baseband waveform moves back ($-\frac{\Delta f}{2}$) and forth ($+\frac{\Delta f}{2}$) on a circle
to represent bit zero and one.

As we all know, a mixer is a non-linear circuit that generates a new frequency from the two input signals.
A mixer produces in the output signal a component of the product of the two inputs.

\begin{equation}
\begin{aligned}
&cos(2\pi f t)cos(2\pi \Delta f t)
\\=&\frac{1}{2}cos(2\pi (f+\Delta f)t) + \frac{1}{2}cos(2\pi (f-\Delta f)t)\label{mixer}
\end{aligned}
\end{equation}

\begin{figure}[htbp]
\centerline{\includegraphics{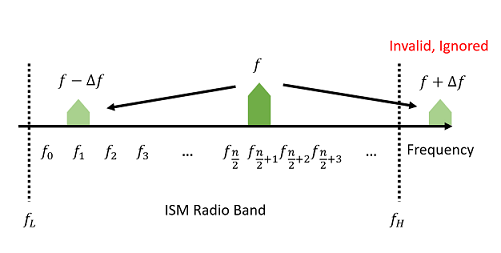}}
\caption{A mixer spreads one frequency into a upper one and a lower one. Only one of them is a valid codeword.}
\label{fsk}
\end{figure}

The mixer therefore spreads the frequency $f$ into two new frequencies $f+\Delta f$ and $f-\Delta f$\cite{mixer}.
To make the demodulation unambiguous, we want only one of these two derived frequencies to be a valid symbol (Fig.~\ref{fsk}).
Fortunately, this can be guaranteed under certain conditions, even if the tag is unaware of the exact frequency carrier symbols.

In Fig.~\ref{fsk}, consider an alphabet of size $n$, where $n$ is an even number.
Adjacent frequencies in the alphabet are equally spaced.
For a particular frequency, without loss of generality, let $k \geq \frac{n}{2}$.
Then for the mixer input $\Delta f$, We want $f_k - \Delta f$ to be a valid codeword, but $f_k + \Delta f$ is not.
We can achieve this by simply setting $\Delta f$ to $f_\frac{n}{2} - f_0$ for all $f_k$.

The invalid frequency generated by the tag can be eliminated with a bandpass filter at the tag or the receiver.
In Bluetooth, $n = 2$, and any frequency components outside the target ISM radio channel is considered as interference and ignored.
That is, as long as $f_1 + \Delta f > f_H$ and $f_0 - \Delta f < f_L$, the demodulator then automatically ignores these unintended components.
For an ISM channel bandwidth $w = f_H - f_L$, as a rule the modulation index $\frac{f_1-f_0}{w}$ is $0.5$,
so it is feasible not to purposely eliminate unintended frequencies from a tag.

Gaussian frequency-shift keying (GFSK) aims to smooth the transitions of the modulated signal.
One significant advantage of this filter is its ability to reduce sideband power and interference with neighboring channels.
Though Gaussian filtering increases the inter symbol interference (ISI),
it does not change the central frequency,
and there is no need to reconfigure frequency shift operations of the tag.

\subsection{Phase Shift Keying}\label{sec_psk}

PSK transmits bits with phase differences.
For M-PSK, the symbols can be expressed as:

\begin{equation}
\begin{aligned}
S(t) = A cos(2\pi (\frac{i}{M} + f t)), \\{\rm where}\ 
i = 0, 1, ..., M-1\label{psk}
\end{aligned}
\end{equation}

\begin{figure}[htbp]
\centerline{\includegraphics{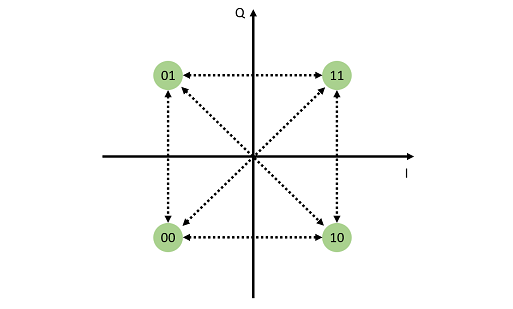}}
\caption{Constellation diagram of the QPSK symbol change trajectory.}
\label{qpsk1}
\end{figure}

4-PSK, also known as QPSK, is a form of M-PSK.
Fig.~\ref{qpsk1} shows that on a constellation diagram, bits are represented by four symmetrical points located around a circle.
In \eqref{psk_delay}, we can transform one to another valid codeword at the tag by delaying the signal in the time domain,
and the value of $k$ corresponds to the tag bits in the alphabet.

\begin{equation}
\begin{aligned}
S'(t) =&S(t - t_T)
\\=&A cos(2\pi (\frac{i}{M} + f (t-t_T)))
\\=&A cos(2\pi \frac{i-k}{M}),
\\{\rm where}\ t_T = \frac{1}{f}\frac{k}{M}\ &{\rm and}\ k = 0, 1, ..., M-1\label{psk_delay}
\end{aligned}
\end{equation}

The presence of 180° phase shifts causes the envelope of the signal to pass through zero instantaneously,
which is reflected in the spectrum as a broadening.
The Offset QPSK modulation method gives a solution that prohibits 180° phase shifts,
as at most one of the in-phase signal and the quadrature signal toggles its polarity at the same symbol boundary
after introducing a half-bit offset between I and Q bit sequences.

\begin{figure}[htbp]
\centerline{\includegraphics{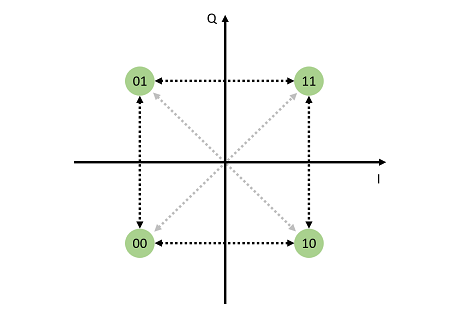}}
\caption{OQPSK avoids 180° phase shifts by introducing a half-bit offset between two orthogonal BPSK signals.}
\label{qpsk2}
\end{figure}

There is a confliction between modifying the OQPSK signal at the tag
and preserving the validity of the symbol sequence,
since based on our design goals discussed in \ref{as_tag},
generating 180° phase shifts in the backscattered signal is unavoidable.
Thankfully a tag is only expected to ensure that most of the symbol sequence is decodeable,
so that the receiver can recognise them.

\begin{figure}[htbp]
\centerline{\includegraphics{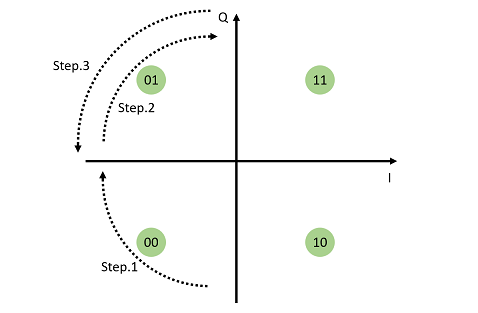}}
\caption{MSK waveform representing consecutive symbols: 00, 01, 01.}
\label{qpsk3}
\end{figure}

A significant proportion of devices today modulates OQPSK signals by minimum-shift keying (MSK) modulation.
MSK uses one sine wave of constant amplitude similar to FSK to modulate the phase,
as shown in Fig.~\ref{qpsk3},
rather than a rectangle from OQPSK symbol switching.
It leads to a more constant envelope than two BPSK signals stacked.

In MSK, the current phase and the upcoming bit determines whether the waveform goes forward or backward.
The reason why two frequencies $+f$ and $-f$ are enough for keeping in phase with OQPSK modulated signal,
is that the polarity of the two orthogonal components changes in turn.
There is no such codeword sequence as ``00, 01, 00''.

Though OQPSK has been implemented with minimum-shift keying,
it is still not appropriate for the tag to do phase shifting in a frequency shift manner.
As we introduced in section \ref{sec_fsk}, a mixer always produces an unexpected frequency component.
This causes interference and ambiguity in phase.
If a tag toggles its mixer input to perform a 90 degrees phase shift,
the expected behaviour is that it will invert the trajectory of the original waveform.
but it results in two different phases in one symbol when non-minimum frequencies are not filtered.
If performing a 180 degrees phase shift,
it is prone to produce consecutive decoding failure and conflict with the minimum frequency.
Therefore, applying time domain delays rather than frequency shifting is the appropriate method for phase shift keying.

\section{Backscatter Protocol Implementation}\label{sec4}

\subsection{Frequency Modulation on Bluetooth}\label{sec41}
In Bluetooth, two available symbols means that
there are two states of the reflection on the tag:
Applying a frequency $\Delta f = f_2 - f_1$ or $0$ to the mixer.
One state represents one tag bit.
In order to keep the structure of most symbols for decodability,
we would like to keep the send rate of tag data down,
i. e. multiple symbols carry one tag bit.

The Bluetooth modules on receivers will decode the signal from a backscatter device in a normal way.
Receivers knows that the tag embeds with bit `0' if the backscattered codeword is the same as the carrier's original codeword,
or bit `1' if it is different.
There is no need to worry about some decoding errors for multiple symbols carrying one bit.
We get a maximum transfer rate of $(1000/n)$Kbps if $n$ symbols carrying one tag bit.

\subsection{Phase Modulation on ZigBee}\label{sec42}

Based on section \ref{sec_psk}, we can introduce at most four different modifications $0, \frac{\pi}{2}, \pi, \frac{3\pi}{2}$ on QPSK or OQPSK modulated signals.
The $\pi$ phase shift will cause decoding errors since a symbol sequence contains simultaneous reversal of polarity is not valid in OQPSK.

Notice that if we only enable $0$ and $\pi$ phase delays,
the modified signal has a good property that the decoded bit sequence still has a continuous inverse taken.
This is obvious while considering that the signal is a stack of two BPSK signals,
meanwhile we have another way to view what would happen if $\pi$ phase shift applied.

\begin{itemize}
\item A codeword `00' (or `11') stands for consecutive zeros (ones), and it is converted to `11'(or `00').
\item A codeword `01' (or `10') stands for a flip in the bit sequence, and it is converted to `10' (or `01').
\end{itemize}

So ideally we would get an exactly reversed decoded bit sequence.
This provides facilitation on decoding and error correction.
In this case it is meaningful to make multiple symbols carrying one tag bit,
not only does it reduce decoding failures caused by corrupting phase continuity,
but sidesteps the difficulty of tag bits recovery due to undesirable inversions on the tag bit boundary.

To take advantage of this concise property,
we now make a convention that only a phase delay of 0 or 180 degrees are allowed at the tag.
And we will utilize this for decoding at the receiver in section \ref{decode}.

\subsection{Decoding Tag Bits at Receivers}\label{decode}

We have described the criteria for generating FSK and OQPSK backscatter signals in section \ref{sec41} and \ref{sec42},
so naturally there are corresponding decoding processes for the receiver.
It is anticipated that the bit error rate (BER) is close to 0 in an ideal noise-free channel.
To verify that the tag bits are correctly decoded in line with expectations,
a simulation was run on Matlab based on the following constraints:

\begin{itemize}
\item The exciters and receivers are connected in a relatively reliable commodity network,
so bit errors in non-backscatter communication are corrected by the software.
In realistic scenarios, backscatter communications are difficult to succeed if non-backscatter ones are no longer stable.
\item We simply assume that if more than half of the carrier bits in which one tag bit locates is flipped,
the receiver treats it as bit `1', otherwise `0'.
\end{itemize}

\begin{figure}[htbp]
\centerline{\includegraphics{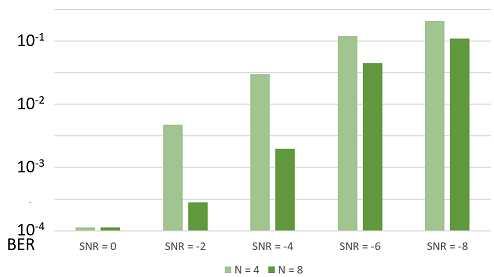}}
\caption{Decoding BER in frequency shift modulation while embedding one tag bit into $N = 4$ or $8$ symbols.}
\label{sim_fsk}
\end{figure}

\begin{figure}[htbp]
\centerline{\includegraphics{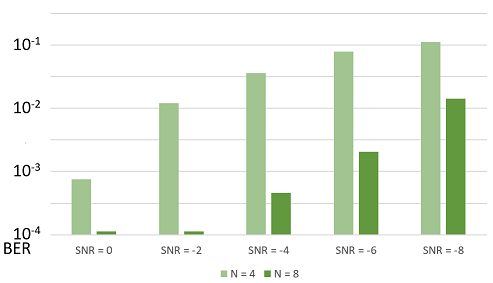}}
\caption{Decoding BER in offset phase shift modulation while embedding one tag bit into $N = 4$ or $8$ symbols.}
\label{sim_oqpsk}
\end{figure}

Note that the signal-to-noise ratio (SNR) baseline is defined on non-backscatter communication, i. e. the carrier wave.
In Fig.~\ref{sim_fsk} and \ref{sim_oqpsk}, adjusting $N$ from 4 to 8 significantly improved the BER performance.
However continuing to increase $N$ will no more make a sharp improve.
For example, when SNR = $-8$ in FSK, changing $N$ from 8 to 16 only reduces the BER from approximately 12\% to 10\%,
as there are already too many bit errors in the carrier signal.
$N = 16$ similarly showed attenuated improvement in offset QPSK.

$N = 8$ performs greater performance improvement in OQPSK than FSK,
partly due to $N = 4$'s difficulty in dealing with destruction of phase continuity
caused by $\pi$ phase delay at the tag.

\section{Conclusion}

The innovation of FreeRider is that its backscatter system works on productive commodity radio,
which is supposed to reduce the collisions and congestions in the target radio band.
At the same time, the hardware requirements were simplified
for a software-configured commodity gateway or terminal can serve as an exciter or coordinator\cite{c0}.
While an exclusive hardware at the exciter is no longer forced,
it complicates the decoding process at the receivers.
Another thing is that although the exciter does not need to produce non-productive (or backscatter-only) communications,
packet length modulation as required by a MAC protocol still requires an coordinator with specific software functionality.
So compared with previous works\cite{passive_wifi, inter_scatter}, it is a compromise plan that does not benefit both the transmitter and the receiver.
In a crowded environment with devices, FreeRider's interference avoidance sheme\cite{c0} may be ineffective,
and therefore may have difficulty truly improving radio band efficiency in production environments.

\section*{Acknowledgment}

This is a final report for the course Parallel and Distributed Computing (COMP.6107P).
I am grateful for teacher Wei Gong's contribution to the course,
and appreciate for teaching assistant Zhaoyuan Xu's guidance on this report.

\end{document}